\documentstyle[12pt,a4]{article}
\global\arraycolsep=2pt %reduces the separation in eqnarrays
\input{epsf}
\begin{document}
%-----------------------------
\makeatletter
\def\fmslash{\@ifnextchar[{\fmsl@sh}{\fmsl@sh[0mu]}}
\def\fmsl@sh[#1]#2{%
  \mathchoice
    {\@fmsl@sh\displaystyle{#1}{#2}}%
    {\@fmsl@sh\textstyle{#1}{#2}}%
    {\@fmsl@sh\scriptstyle{#1}{#2}}%
    {\@fmsl@sh\scriptscriptstyle{#1}{#2}}}
\def\@fmsl@sh#1#2#3{\m@th\ooalign{$\hfil#1\mkern#2/\hfil$\crcr$#1#3$}}
\makeatother
%--------------------------------
%---------------- CERN Title page <---------------------------
%\thispagestyle{empty}
\begin{titlepage}

\begin{flushright}
%hep-ph/0003283 \\
{\bf LMU 17/99}\\
\today
\end{flushright}

\vspace{0.3cm}
\boldmath
\begin{center}
  \Large \bf{ The weak angle $\gamma$ from one--particle inclusive CP
    asymmetries in the $B^0_s$ system.}
\end{center}
\unboldmath
\vspace{0.8cm}

\begin{center}
  {\large Xavier Calmet} \\
  {\sl Ludwigs-Maximilians-Universit{\"a}t, Sektion Physik,
    Theresienstr. 37, \\ D-80333 M{\"u}nchen, Germany}
\end{center}

\vspace{\fill}

\begin{abstract}
\noindent 
We compute CP asymmetries of one-particle inclusive $B^0_s \to D_s X$
decay rates.  We find that the weak angle $\gamma$
could in principle be extracted from these asymmetries although they
are very small.  Taking the decay width difference into account, we
find that the time integrated CP asymmetries are of the order ${\cal
  A}=1.4\ 10^{-4} \sin(\gamma)$. Some large
uncertainties remain due to the lack of experimental data.
\end{abstract}
{to appear in Physical Review D.}
\end{titlepage}

% ----------------------------------------------------------
%%%%%%%%%%%%%%%%%%%%%%%%%%%%%%%%%%%%%%%%%%%%%%%%%%%%%%%%%%%%%%%%%%%%%
\section{Introduction}
Some time ago it was proposed by Beneke, Buchalla and Dunietz
\cite{BBD} to study mixing induced CP violation in $B$ mesons by using
asymmetries of inclusive decay rates. This method although
theoretically clean suffers from an experimental difficulty since it
supposes a completely inclusive measurement. In order to solve this
problem, we propose to study CP asymmetries of one-particle inclusive
decays. In this paper, we shall consider CP asymmetries in the $B^0_s$
system. This can also be done in the $B_{d}$ system \cite{CMS2}.  We shall
study time integrated and time dependent CP asymmetries of
one-particle inclusive $B^0_s \to D_s X$ decays.  The technique involved is
nevertheless theoretically not as clean as the one used in \cite{BBD}
for inclusive decays.  Nevertheless, the measurement of CP asymmetries
of one-particle inclusive decay widths would be clean.

We shall not neglect the decay width difference in the $B^0_s$ system,
since it is sizable although recent calculations \cite{BBG} show that
the decay width difference in that system could be smaller than
previously expected. A method \cite{calmet1} was recently proposed to
calculate the decay rates appearing in these asymmetries. The
predicted decay rates are compatible with current experimental
knowledge and we can therefore be confident that the results we obtain
for the CP asymmetries will be meaningful.

We shall concentrate on asymmetries involving one-particle inclusive
decays of a $B^0_s$ meson. We start by introducing our notations, in
section \ref{formalism} we shall develop the formalism needed 
for time dependent CP asymmetries in one-particle CP asymmetries. 
We then present and discuss our results and conclude.
%%%%%%%%%%%%%%%%%%%%%%%%%%%%%%%%%%%%%%%%%%%%%%%%%%%%%%%%%%%%%%%%%%%%%
\section{Definitions}
In this section we shall introduce our notations. We basically use
notations similar to those introduced in \cite{BBD}.  The proper time
evolution of an initial pure $B^0_s$ or $\bar B^0_s $ reads
\begin{eqnarray}
\mid B^0_{s \ {phys}}(t) \rangle &=&g_+ (t) \mid B^0_s \rangle -
       \frac{q}{p}g_- (t) \mid \bar B^0 _s 
\rangle  \\ 
\mid \bar B^0 _{s \ phys}(t) \rangle &=& -\frac{p}{q}g_- (t)\mid B^0_s 
   \rangle + g_+ (t)  \mid \bar B^0 _s \rangle,  \nonumber
\end{eqnarray}
where the time dependent functions
\begin{eqnarray}
g_+ (t)&=& e^{-iMt-\frac{1}{2} \Gamma t}\left[\cosh 
\frac{\Delta{\Gamma}t}{4} \cos\frac{\Delta{M}t}{2} + i \sinh 
\frac{\Delta{\Gamma}t}{4} \sin\frac{\Delta{M}t}{2} \right]
\\ \nonumber
\end{eqnarray}
and
\begin{eqnarray}
g_- (t)&=&  e^{-iMt-\frac{1}{2} \Gamma t}\left[\sinh 
\frac{\Delta{\Gamma}t}{4} \cos\frac{\Delta{M}t}{2} + i \cosh 
\frac{\Delta{\Gamma}t}{4} \sin\frac{\Delta{M}t}{2} \right]
\end{eqnarray}
describe the particle anti-particle mixing. 
The mass difference $\Delta{M}$ and the width difference 
$\Delta{\Gamma}$  between the neutral $B^0_s$ mesons are given by
\begin{eqnarray}
\Delta{M}&=&M_H-M_L \\ \nonumber
\Delta{\Gamma}&=&\Gamma_H-\Gamma_L. 
\end{eqnarray}
The off diagonal term of the mass-mixing matrix is given by 
$M_{12}+i\Gamma_{12}$. We define the quantity
\begin{eqnarray} \label{pq}
\frac{q}{p}=\frac{\Delta{M}-i/2\Delta{\Gamma}}{2\left (M_{12}-
i/2\Gamma_{12} \right 
)}=\frac{M_{12}^*}{|M_{12}|}\left(1-\frac{1}{2}a\right), \ \ \ \ \ 
a={\rm 
Im}\frac{\Gamma_{12}}{M_{12}}.
\end{eqnarray}
The second expression for $q/p$ in equation (\ref{pq}) is valid to first
order in the 
small quantity $\Gamma_{12}/M_{12}={\cal O}(m_b^2/m_t^2)$.
%%%%%%%%%%%%%%%%%%%%%%%%%%%%%%%%%%%%%%%%%%%%%%%%%%%%%%%%%%%%%%%%%%%
\section{CP asymmetries in one-particle inclusive decays} \label{formalism}

In this section, we shall consider CP asymmetries of rates of
one-particle inclusive $B^0_s \to D_s X$ decays. The time
dependent CP asymmetry is defined by
\begin{eqnarray}
{\cal A}_{CP} (t)=\frac{ \Gamma (B^0_s(t) \rightarrow D_s X) - 
\Gamma(\bar B^0 _s(t)\to  \bar D_s \bar X)}{\Gamma(B^0_s(t) \rightarrow 
 D_s X) + \Gamma(\bar B^0_s(t)\to  \bar D_s \bar X)}. 
\end{eqnarray}
In the following, we shall neglect all effects due to direct CP 
violation. The time dependent decay widths therefore read
\begin{eqnarray}
\Gamma(B^0_s(t) \rightarrow D_s X) = \arrowvert g_+(t) 
\arrowvert ^2 \Gamma(B^0_s \rightarrow D_s X) + \arrowvert 
\frac{q}{p}g _-(t) \arrowvert^2 \Gamma(B^0_s \rightarrow \bar D_s X )
\\
- 2 {\rm Re} \ \left (  g_+^{\star}(t)  \frac{q}{p}g _-(t) T^{ B^0_s \ 
\bar B^0 _s }_{ D_s } \right ) \nonumber
\end{eqnarray}
and
\begin{eqnarray}
\Gamma(\bar B^0_s(t) \to \bar D_s \bar{X}) = \arrowvert g_+(t) 
\arrowvert ^2 \Gamma(B^0_s \rightarrow D_s X) + \arrowvert 
\frac{p}{q}g _-(t) \arrowvert^2 \Gamma(B^0_s \rightarrow \bar D_s X)  
\\ -
2 {\rm Re} \ \left (  g_+^{\star}(t)  \frac{p}{q}g _-(t) T^{\bar B^0_s \ 
{B^0_s}}_{\bar D_s} \right ). \nonumber
\end{eqnarray}
The decay widths appearing in these formulas were 
 calculated in \cite{calmet1}. It remains to compute the transition 
 matrix elements of $\Delta{B^0_s}=2$ given by
 \begin{eqnarray} \label{Tw}
T^{B^0_s \ \bar B^0 _s }_{ D_s } &=& \frac{1}{2 m_{B^0_s}} \int
d^4x \int d\phi_{D_s}\sum_{X} (2 \pi)^4 \delta^4(P_{B^0_s}-P_{D_s}-P_X) \\ 
\nonumber 
&&
\langle B_s^0 \arrowvert H_{eff}(x) \arrowvert D_s X \rangle 
\langle D_s X  \arrowvert H_{eff}^\dagger(0) \arrowvert 
\bar B_s^0 \rangle
\end{eqnarray}
and
\begin{eqnarray}\label{Tr}
T^{\bar B^0_s \ {B^0}_s }_{\bar D_s} &=& \frac{1}{2 m_{B^0_s}} \int
d^4x \int d\phi_{\bar D_s} \sum_{\bar X} (2 \pi)^4 \delta^4(P_{B^0_s}-P_{\bar D_s}-P_{\bar X}) 
\\ \nonumber &&
\langle \bar B_s^0 \arrowvert H_{eff}(x) \arrowvert \bar D_s \bar X \rangle 
\langle \bar D_s \bar X \arrowvert H_{eff}^\dagger(0) \arrowvert B_s^0 \rangle,
\end{eqnarray}
where $d\phi_{D_s}$ is the phase space of the $D_s$ meson. The part of
the effective Hamiltonian which is relevant for this work is given by
\begin{eqnarray}\label{effHam}
  H_{eff}&=&\frac{G_F}{\sqrt{2}} V_{cb}^* V_{cs} \left ( C_1 (\bar b c)_{V-A}
    (\bar c s)_{V-A}
  + C_2 (\bar b T^a c)_{V-A} (\bar c T^a s)_{V-A} \right) \\ && +
\frac{G_F}{\sqrt{2}} V_{cb}^* V_{us} \left ( C_1 (\bar b c)_{V-A}
    (\bar u s)_{V-A}
  + C_2 (\bar b T^a c)_{V-A} (\bar u T^a s)_{V-A} \right)
 \nonumber \\ && +
\frac{G_F}{\sqrt{2}} V_{ub}^* V_{cs} \left ( C_1 (\bar b u)_{V-A}
    (\bar c s)_{V-A}
  + C_2 (\bar b T^a u)_{V-A} (\bar c T^a s)_{V-A} \right)\nonumber 
\end{eqnarray}
where $(\bar q_1 q_2)_{V-A}$ stands for $(\bar q_1
\gamma^\mu(1-\gamma_5) q_2)$, the $T^a$ matrices are the $SU(3)_C$
Gell-Mann matrices and $C_1$ and $C_2$ are the Wilson coefficients.
The matrix elements $T^{B^0_s \ \bar B^0 _s }_{ D_s }$ and $T^{\bar
  B^0_s \ {B^0}_s }_{\bar D_s}$ can be parametrized in the same way as
the wrong charm decay in \cite{CMS}. Applying the Fierz transformation we
can rewrite the operators into the following form $(\bar b s)_{V-A}
(\bar c c)_{V-A}$, $(\bar b s)_{V-A} (\bar u c)_{V-A}$ and $(\bar b
s)_{V-A} (\bar c u)_{V-A}$. Using the large $N_C$ limit, we can
factorize $T^{\bar B^0_s \ B^0_s }_{\bar D_s}$
\begin{eqnarray} 
T^{B^0_s \ \bar B^0_s  }_{\bar D_s}(M^{2})&=& \sum_q \frac{1}{2 m_{B^0_s}}
   \frac{G_F^2}{2} 
|C_1|^2  V_{cb}^*  V_{qs}  V_{qb}^*  V_{cs} \\ && \nonumber  
\int  d^4Q\,  K^{\mu \nu}(p_{B^0_s},M,Q)
\int d\phi_{D_s} P^{(q)}_{\mu \nu}(Q) 
\end{eqnarray}
with
\begin{eqnarray}
K_{\mu \nu} (p_{B^0_s},M,Q) &=& \sum_X (2 \pi)^4 \delta^4 (p_{B^0_s} - p_X - Q)
\\ \nonumber &&
\langle \bar B^0_s(p_{B^0_s}) | (\bar{b} \gamma_\mu (1-\gamma_5) s)|  X  \rangle
\langle X | (\bar{b} \gamma_\nu (1-\gamma_5) s)| B^0_s(p_{B^0_s})  \rangle  
\end{eqnarray}
and 
\begin{eqnarray}
P^{(q)}_{\mu \nu} (Q) &=&  \sum_{X'} (2 \pi)^4 \delta^4 (Q - p_{D_s} - p_{X'}) 
\\ \nonumber && 
\langle 0 |  (\bar{c} \gamma_\mu (1-\gamma_5) q) 
          |\bar  D_s(p_{\bar D_s}) X'  \rangle 
\langle \bar  D_s(p_{\bar D_s}) X'  |  (\bar{q} \gamma_\nu (1-\gamma_5) c) 
          | 0 \rangle.
\end{eqnarray}
We get a similar expression for $T^{ B^0_s \ \bar B^0_s }_{ D_s}(M^{2})$.
The nature of the quark $q$ depends on the operator under
consideration.  The tensor $K_{\mu \nu} (p_{B^0_s},M,Q)$ is fully
inclusive and can be parametrized using the decay constant $f_{B^0_{s}}$
of the $B^0_s$ meson and $P^{(q)}_{\mu \nu} (Q)$ is very similar to the
expression we had encountered in the calculation of the wrong charm
decay width \cite{CMS}. It involves a projection on a state containing
a $D_s$ meson. A priori we do not know how to calculate that kind of
matrix element, a solution is to contract the spinor indices as in the
parton calculation and to multiply this tensor by a channel dependent
form factor $f_q$ which has to be fitted. We obtain
\begin{equation}
  P^{(q)}_{\!\mu\nu}(p_D,\!Q) \!=\! 2\pi \,
    \delta \! \left( \left( Q \!-\! p_D \right)^2 \!\!-\! m_q^2 \right)
    {\rm Tr} \Big\{ \fmslash{p}_D \, \gamma{_\mu}
      \left( \fmslash{Q} \!-\! \fmslash{p}_D \right) \gamma{_\nu} \!\Big\}
    f_q.
\end{equation}
We can reproduce the inclusive case by setting $f_q=1$. 
%%%%%%%%%%%%%%%%%%%%%%%%%%%%%%%%%%%%%%%%%%%%%%%%%%%%%%%%%%%%%
\subsection{CP asymmetry $\Gamma(B^0_s(t) \to D^{+}_{s} X)$ vs.
  $\Gamma( \bar B^0_s(t) \to D^{-}_{s} \bar X) $}
The CP asymmetry reads
\begin{eqnarray}
{\cal A}(t)&=&\frac{\Gamma(B^0_s(t) \to D^{+}_{s} X) -
\Gamma( \bar B^0_s(t) \to D^{-}_{s} \bar X)}{\Gamma(B^0_s(t) \to D^{+}_{s} 
X) + \Gamma( \bar B^0_s(t) \to D^{-}_{s} \bar X)}.
\end{eqnarray}
We have two quark transitions contributing to $T^{B^0_s \ \bar B^0 _s }_{
  D^+_s }$, namely $b \to c \bar c s$ interfering with itself and $b
\to u \bar c s$ interfering with $b\to c \bar u s$.  These matrix
elements can be modeled in the same way as it was proposed in
\cite{calmet1}. It corresponds to a rescaling of the parton
calculation. The complication due to isospin symmetry which led to
the introduction of strong phases in \cite{CMS2} is not present in the
$B^0_s$ system. We can use the same parameterization for the
$\Delta{B^0_{s}}=2$ as the one we have used for the wrong charm decay
widths \cite{calmet1}.  The parton calculation was performed in
\cite{BBD}. We obtain
 \begin{eqnarray} \label{TBD}
T^{B^0_s \ \bar B^0_s}_{D^+_s}&=&
- \frac{G_F^2 m_b^2}{24 \pi M_{B^0_s}} 
f_{B^0_s}^2 M_{B^0_s}^2 V_{cb} V^*_{us} V_{ub} V^*_{cs}  (1-z)^2
 \\ && \cdot
\bigg 
( C_1^2 (1-z) B  + C_1^2 (1+2 z)  \frac{M_{B^0_s}^2}{m_b^2} 2 B^0_S \bigg )
{\cal G} 
\nonumber \\ &&
-\frac{G_F^2 m_b^2 f_{B^0_s}^2}{24 \pi M_{B^0_s}} 
M_{B^0_s}^2
  (V^*_{cs} V_{cb})^2 
  \sqrt{1-4 z} \nonumber \\ && \cdot
  \bigg[  (1-4 z)C_1^2 B 
  +(1+2 z) C_{1}^2 \frac{M_{B^0_s}^2}{m_b^2} 2 B_S \bigg ] {\cal F},
\nonumber 
\end{eqnarray}
in the leading order of the short distance expansion and where ${\cal
  F}$ and ${\cal G}$ are decay channel dependent non-perturbative form
factors. We shall define ${\cal F}$ and ${\cal G}$ later. The
parameters $B$ and $B_S$ are the bag factors and $z$ is defined by
\begin{eqnarray}
  z=\frac{m_c^2}{m_b^2}.
\end{eqnarray}
We have
neglected the penguin operators. We also have two
operators contributing to $T^{\bar B^0_s \ {B^0}_s }_{ D^-_s}$, as in
the previous case $b \to c \bar c s$ interfering with itself and $b\to
c \bar u s$ interfering with $b \to u \bar c s$. In the leading order
of the short distance expansion we then have
 \begin{eqnarray} \label{TBDbar}
T^{\bar B^0_s \ B^0_s}_{D^-_s}&=&- \frac{G_F^2 m_b^2}{24 \pi M_{B^0_s}} 
f_{B^0_s}^2 M_{B^0_s}^2 V^*_{cb} V_{us} V^*_{ub} V_{cs}  (1-z)^2
 \\ && \cdot
\bigg 
( C_1^2 (1-z) B  + C_1^2 (1+2 z)  \frac{M_{B^0_s}^2}{m_b^2} 2 B_S \bigg )
{\cal G} 
\nonumber \\ && -\frac{G_F^2 m_b^2 f_{B^0_s}^2}{24 \pi M_{B^0_s}}M_{B^0_s}^2
  (V^*_{cb} V_{cs})^2 
  \sqrt{1-4 z} \nonumber \\ && \cdot
  \bigg[  (1-4 z)C_1^2 B 
  +(1+2 z)C_1^2 \frac{M_{B^0_s}^2}{m_b^2} 2 B_S \bigg ]  {\cal F},
  \nonumber 
  \end{eqnarray}
  where ${\cal F}$ and ${\cal G}$ are decay channel dependent
  non-perturbative form factors. The function ${\cal F}$ was given in
  \cite{calmet1}. We had
\begin{eqnarray}
 {\cal  F}^{B_s^0 D_s^+}={\cal F}^{B_s^0 D_s^-}=4 f,
\end{eqnarray}
where $f=0.121$ was fitted in \cite{CMS}. The value of this parameter
could be slightly different in the $B_s^0$ case and will eventually
have to be extracted from experiment. It can be extracted easily from the
decay channel $B^0_s \to D^+_s X$. For the decays involving a $D_s^*$
we have
\begin{eqnarray}
  {\cal F}^{B_s^0 D_s^{*+}}={\cal F}^{B_s^0 D_s^{*-}}=3 f.
\end{eqnarray}
We shall make the very same assumptions to model ${\cal G}$ as we made
to model $ {\cal F}$. To do so we have to consider the contribution
of $b \to u \bar c s$ to the decay rates of the form $\Gamma(B^0_s \to
D_s X)$ which we shall denote by $\Gamma(B^0_s \to D_s X)^{b \to u
  \bar c s}$. Using the same model as the one we used for the wrong
charm case in \cite{CMS}, we have
\begin{eqnarray}
\Gamma(B^0_s \to D^{* +}_s X)^{b \to u \bar c
  s} =3
\Gamma(B^0_s \to D^{+}_s X)^{b \to u \bar c s}_{dir},
\end{eqnarray} 
where $\Gamma(B^0_s \to D^{+}_s X)^{b \to u \bar c
  s}_{dir}=\Gamma_{dir}^{b \to u \bar c s}$ is the direct contribution
to this decay over the current $b \to u \bar c s$. If we take into
account the contribution from the decay over a $D^{*+}_s$, we obtain
\begin{eqnarray}
\Gamma(B^0_s \to D^{+}_s X)^{b \to u \bar c
  s} &=& 4 \Gamma_{dir}^{b \to u \bar c
  s}
\end{eqnarray}
since a $D^{*+}_s$ always decays into a $D^{+}_s$.  We assume that the
direct contribution $\Gamma_{dir}^{b \to u \bar c s}$ to $\Gamma(B^0_s
\to D^{+}_s X)^{b \to u \bar c s}$ is equal to the direct contribution
to $\Gamma(B^0 \to D^{+} X)^{b \to u \bar c s}$ which is certainly the
case in the limit of the heavy quark symmetry. We assume that every
$c$ quark eventually hadronizes into a $D$ meson. Spin counting and
isospin symmetry in the decays $B \to D^* X$ and $B \to D_{dir} X$
through the transition $b \to u \bar c s$ allow us to deduce easily
that $\Gamma_{dir}$ is equal to $1/8$ times the result of the parton
calculation. We then obtain the following non-perturbative form factors
\begin{eqnarray}
{\cal G}^{B_s^0 D_s^+}={\cal G}^{B_s^0 D_s^-}=1/2
\end{eqnarray}
and
\begin{eqnarray}
{\cal  G}^{B^0_s D_s^{*+}}={\cal G}^{B^0_s D_s^{*-}} =3/8.
\end{eqnarray}
  In the $B^0_s$ system, we can use the following approximations
\begin{eqnarray}
   a=0  \ \ {\rm and} \ \ \frac{M_{12}^*}{|M_{12}|}= \frac{M_{12}}{|M_{12}|}=1.
  \end{eqnarray}
  These are very good approximations, theory predicts 
  $a<10^{-3}$, we then neglect the weak phase in 
  $\frac{M_{12}^*}{|M_{12}|}$.  We see
  that the only weak phases appearing are $e^{i \gamma}$ in equation
  (\ref{TBDbar}) and $e^{-i \gamma}$ in equation (\ref{TBD}). 
   We can therefore rewrite $T^{B^0_s \ \bar B^0_s}_{D_s}$
  and $T^{\bar B^0 \ B^0}_{\bar D_s} $ as
\begin{eqnarray}
  T^{B^0_s \ \bar B^0_s}_{D^+_s}=n_1+ e^{- i \gamma} n_2
  \end{eqnarray}
and
\begin{eqnarray}
  T^{\bar B^0_s \ B^0_s}_{ D^-_s}=n_1 + e^{i  \gamma} n_2, 
 \end{eqnarray}
 where $n_1$ corresponds to the contribution from $b \to c \bar c s$
 interfering with it-self and $e^{i \gamma} n_2$ corresponds to the
 contribution from $b \to c \bar u s$ interfering with $b \to u \bar c
 s$.
%%%%%%%%%%%%%%%%%%%%%%%%%%%%%%%%%%%%%%%%%%%%%%%%%%%%%%%%%%%%%%%%%%%%%%%%5
Inserting the formulas for the time dependent decay rates in the
 definition for the CP asymmetry, we obtain
\begin{eqnarray}
{\cal A}_{CP}(t)=\frac{- 2 n_2 \sin\left(\Delta{M} t \right)
  \sin(\gamma)}{M_1(t)
  2 \Gamma(B^0_s \to D^+_s X)+M_2(t) 2 \Gamma(B^0_s \to D^-_s X) + M_3(t) },
\end{eqnarray}
where $M_1(t), M_2(t)$ and $M_3(t)$ are given by
\begin{eqnarray}\label{Eq1}
  M_1(t)=\cos^2\left(\frac{\Delta{M}t}{2}  \right)+
  \sinh^2\left (\frac{\Delta{\Gamma} t}{4}   \right),
 \end{eqnarray}
 \begin{eqnarray}\label{Eq2}
   M_2(t)=\sin^2\left(\frac{\Delta{M}t}{2}  \right)
   +\sinh^2\left (\frac{\Delta{\Gamma} t}{4}   \right)
 \end{eqnarray}
 and
 \begin{eqnarray} \label{Eq3}
   M_3(t)=   -(2 n_{1} +2 n_2) \sinh\left(\frac{\Delta{\Gamma}t}{2}\right).
 \end{eqnarray}
The time integrated CP asymmetry is the given by
  \begin{eqnarray}
{\cal A}_{CP}=\frac{- 2 n_2  \frac{x}{1+x^2} \sin(\gamma) }{
 N_1(x,y)\Gamma(B^0_s \to  D^+_s X)+ 
 N_2(x,y) \Gamma(B^0_s \to D^-_s X) +  N_3(x,y)},
 \end{eqnarray}
 where $N_1(x,y)$, $N_2(x,y)$ and $N_3(x,y)$ are given by
  \begin{eqnarray} \label{Eq4}
   N_1(x,y)= \frac{2+x^2}{1+x^2}+  \frac{ y^2}{4-y^2},
    \end{eqnarray}
    \begin{eqnarray}\label{Eq5}
   N_2(x,y)= \frac{x^2}{(1+x^2)}+  \frac{y^2}{(4-y^2)}
    \end{eqnarray}
    and
     \begin{eqnarray}\label{Eq6}
   N_3(x,y)= - \frac{2 y}{4-y^2}\left (2 n_1 +2 n_2 \cos(\gamma)\right),
    \end{eqnarray} 
    where we have introduced the parameters $x= \Delta{M}/\Gamma$
    and $y= \Delta{\Gamma}/\Gamma$ which can be measured. At the
    present time, there is only a lower bound for the parameter $x$ in
    the $B^0_s$ system namely $|x|>14$ \cite{PDG}. For numerical
    calculations, we shall use $|x|=20$ \cite{fritzsch}. It is also not
    clear what the actual value of $y$ is, for numerical calculations,
    we shall use the value computed in \cite{BBG}, $|y|= 0.054$. We set
    $B_S=B=1$, $m_{b}= 4.8 \ {\rm GeV}$, $m_{B^0_s}= 5.3693 \ {\rm GeV}$ , 
    $m_{c}= 1.4 \ {\rm GeV}$, $m_{D_s}= 1.9685 \ {\rm GeV}$, 
    $f_{B^0_s}=210 \ {\rm MeV}$ and $C_{1}=1$. We obtain numerically
    \begin{eqnarray}
    n_{1}= -4.931 \ \%, \\
    n_{2}= -9.526 \ 10^{-2} \ \% ,\nonumber
    \end{eqnarray}
    where $n_{1}$ and $n_{2}$ are normalized to the decay width of 
    the $B^0_{s}$ meson, using $\tau_{B^0_s}=1.54 \ 10^{-12} \ {\rm s}$. 
    The decay width needed were computed in \cite{calmet1}, we had
    \begin{eqnarray}
    \Gamma(B^0_{s} \to D^{+}_s X)= 3.3 \% \\
     \Gamma(B^0_{s} \to D^{-}_s X)= 64.9 \%.  \nonumber
    \end{eqnarray}
   The time integrated CP asymmetry is then given by
   \begin{eqnarray}  \label{EqA1}
   {\cal A}_{CP}=1.4 \ 10^{-4} \sin(\gamma),
\end{eqnarray}
neglecting the term proportional to $\cos(\gamma)$ in the denominator.

\subsection{CP asymmetry $\Gamma(B^0_s(t) \to D^{-}_{s} X)$ vs. 
$\Gamma( \bar B^0_s(t) \to D^{+}_{s} \bar X) $}
In this case the CP asymmetry reads
\begin{eqnarray}
{\cal A}(t)&=&\frac{\Gamma(B^0_s(t) \to D^{-}_{s} X) -
\Gamma( \bar B^0_s(t) \to D^{+}_{s} \bar X)}{\Gamma(B^0_s(t) \to D^{-}_{s} 
X) + \Gamma( \bar B^0_s(t) \to D^{+}_{s} \bar X)}.
\end{eqnarray}
We need to parameterize $T^{B^0_s \ \bar B^0_s}_{D^-_s}$ and 
$T^{\bar B^0_s \ B^0_s}_{D^+_s}$. They are given by
\begin{eqnarray}
T^{B^0_s \ \bar B^0_s}_{D^-_s}&=& \left (T^{\bar B^0_s \ B^0_s}_{D^-_s} 
\right)^{\dagger}=n_{1} +e^{-i \gamma} n_{2}\\
T^{\bar B^0_s \ B^0_s}_{D^+_s}&=& \left (T^{B^0_s \ \bar 
B^0_s}_{D^+_s} \right)^{\dagger}=n_1+e^{i \gamma} n_{2}. \nonumber
\end{eqnarray}
The form factors appearing in the transition matrix element were given 
previously.
The time dependent CP asymmetry then reads
\begin{eqnarray}
{\cal A}_{CP}(t)=\frac{-2 n_2 \sin\left(\Delta{M} t \right) 
\sin(\gamma)}{M_1(t)
  2 \Gamma(B^0_s \to D^-_s X)+M_2(t) 2 \Gamma(B^0_s \to D^+_s X) + M_3(t) },
\end{eqnarray}
where $M_1(t)$, $M_2(t)$ and $M_3(t)$ were defined in equations (\ref{Eq1}), 
(\ref{Eq2}) and (\ref{Eq3}).
The time integrated CP asymmetry is then given by
  \begin{eqnarray}
{\cal A}_{CP}=\frac{ -2 n_2 \frac{x}{1+x^2} \sin(\gamma) }{
 N_1(x,y)\Gamma(B^0_s \to  D^-_s X)+ 
 N_2(x,y) \Gamma(B^0_s \to D^+_s X) +  N_3(x,y)},
\nonumber \\ &&
 \end{eqnarray}
 where $N_1(x,y)$, $N_2(x,y)$ and $N_3(x,y)$ are given in equations
 (\ref{Eq4}), (\ref{Eq5}) and (\ref{Eq6}). The time integrated CP
 asymmetry is numerically given by
  \begin{eqnarray}
   {\cal A}_{CP}=1.4 \ 10^{-4} \sin(\gamma),
\end{eqnarray}
neglecting the term proportional to $\cos(\gamma)$ 
in the denominator.
\subsection{CP asymmetry $\Gamma(B^0_s(t) \to D^{*}_{s} X)$ vs. 
$\Gamma( \bar B^0_s(t) \to \bar D^{*}_{s} \bar X) $}
We can without any difficulty deduce from the preceding sections the 
formulas for the CP asymmetries in the one-particle inclusive decays  
of the form $\Gamma(B^0_s(t) \to D^{*}_{s} X)$ versus 
$\Gamma( \bar B^0_s(t) \to \bar D^{*}_{s} \bar X) $.
We have 
\begin{eqnarray}
    n_{1}= -3.698 \ \% \\
    n_{2}= -7.145 \ 10^{-2} \ \% ,\nonumber
    \end{eqnarray}
    which are normalized to the decay width of the $B^0_{s}$ meson.
    The decay widths needed were computed in \cite{calmet1}, we had
    \begin{eqnarray}
    \Gamma(B^0_{s} \to D^{*+}_s X)= 2.5 \ \% \\
     \Gamma(B^0_{s} \to D^{*-}_s X)= 49.6 \ \%.  \nonumber
 \end{eqnarray}
The next CP asymmetry we shall consider is defined by
\begin{eqnarray}
{\cal A}(t)&=&\frac{\Gamma(B^0_s(t) \to D^{*-}_{s} X) -
\Gamma( \bar B^0_s(t) \to D^{*+}_{s} \bar X)}{\Gamma(B^0_s(t) \to D^{*-}_{s} 
X) + \Gamma( \bar B^0_s(t) \to D^{*+}_{s} \bar X)}.
\end{eqnarray}
The time integrated CP asymmetry is then given by
   \begin{eqnarray} 
   {\cal A}_{CP}=1.4 \ 10^{-4} \sin(\gamma),
\end{eqnarray}
neglecting the term proportional to $\cos(\gamma)$ 
in the denominator.
For the asymmetry defined by
\begin{eqnarray}
{\cal A}(t)&=&\frac{\Gamma(B^0_s(t) \to D^{*-}_{s} X) -
\Gamma( \bar B^0_s(t) \to D^{*+}_{s} \bar X)}{\Gamma(B^0_s(t) \to D^{*-}_{s} 
X) + \Gamma( \bar B^0_s(t) \to D^{*+}_{s} \bar X)},
\end{eqnarray}
we obtain
\begin{eqnarray} \label{EqA2}
   {\cal A}_{CP}=1.4 \ 10^{-4} \sin(\gamma),
\end{eqnarray}
neglecting the term proportional to $\cos(\gamma)$ 
in the denominator.
%%%%%%%%%%%%%%%%% Results %%%%%%%%%%%%%%%
\section{Discussion of the results}
We see that in principle we can extract information on $\sin(\gamma)$
from the asymmetries calculated in the previous section. But they are
small, nevertheless the decay widths involved are sizable.  It has
been proposed to extract information on $\sin(\gamma)$ from exclusive
decays (see e.g. \cite{Dunietz}) but the decay widths involved are
very small, typically of the order $10^{-4}$ and one would have to
deal with strong phases which would make the extraction of
$\sin(\gamma)$ even more difficult. It could therefore be worth to try
to extract $\sin(\gamma)$ from one-particle inclusive decays in the
$B^0_s$ system. If the present method is chosen to extract
$\sin(\gamma)$, it would be interesting to test its precision. This
could be done by comparing the results obtained for $\sin(2\beta)$ in
one-particle inclusive CP asymmetries in the $B_d$ system \cite{CMS2}
with some more conventional extraction technique like the ``gold-plated''
$B \to J/\Psi K_S$, although one-particle inclusive CP asymmetries in
the $B_d$ system are theoretically not as clean as the ones in the
$B^0_s$ system due to the presence of strong phases \cite{CMS2}.

We still have some large uncertainties, some of them due to the
method. The corrections to the decay widths could be fairly large, in
the worth case of the order of $30 \%$. But remember that the decay
widths calculated in \cite{calmet1} are compatible with current
experimental knowledge. On the other hand, we have large experimental
uncertainties in the values of $x$, $y$ and $f_{B^0_{s}}$.

Time dependent CP asymmetries could also allow to extract 
$\sin(\gamma)$, but it is not yet clear if the oscillations can be 
resolved. Predictions depend on the decay width difference in the 
$B^0_s$ system and it has not yet been possible to measure this quantity.   

A way to improve the magnitude of the CP asymmetries would be
to do an anti-lepton tagging. The decay widths in the
denominator are partially responsible for the low magnitude of the
asymmetries, this is particularly true for the CP asymmetries in the
$B^0_{d}$ system considered in \cite{CMS2}. Thus, doing an anti-lepton
tagging, would slightly improve their magnitude. We would then consider
asymmetries of the type
\begin{eqnarray}
{\cal A}(t)&=&\frac{\Gamma(B^0_s(t) \to D_{s} X)_{\rm NL} -
\Gamma( \bar B^0_s(t) \to \bar D_{s} \bar X)_{\rm NL}}
{\Gamma(B^0_s(t) \to D_{s} 
X)_{\rm NL} + \Gamma( \bar B^0_s(t) \to \bar D_{s} \bar X)_{\rm NL}},
\end{eqnarray}
where ${\rm NL}$ stands for non-leptonic.  In the $B^0_s$ system, we
would have time integrated CP asymmetries of the order $2 \ 1 0^{-4}
\sin(\gamma)$. The factor $x=20$ is responsible for the low magnitude
of the asymmetries. In the $B_d$ system, this effect would obviously
be larger.
%%%%%%%%%%%%%%%%% Conclusion %%%%%%%%%%%%%%%
\section{Conclusion}
We have discussed CP asymmetries in one-particle inclusive $B^0_{s} \to
D_{s} X$ decays. The asymmetries are small but would allow to extract 
$\sin(\gamma)$ which is known to be difficult. So any new
method is probably welcome. It has the advantage, in comparison to CP
asymmetries of exclusive decays, to have some large decay widths and in
comparison to  CP
asymmetries of inclusive decays, of being experimentally clean.
%%%%%%%%%%%%%%%%% Acknowledgements %%%%%%%%%%%%%%%
\section*{Acknowledgements}
The author is grateful to Z.Z. Xing for long discussions on
$B^0_{s}$ physics and CP violation in that system and for his critical
reading of this manuscript. He would also like to thank A. Leike
for his useful comments.
%%%%%%%%%%%%%%%%%%%%%% Biblio %%%%%%%%%%%%%%%%%%%%%%%%%%%%%%%%%%%%%%%%%%%%%

\end{document}